# Cell mapping description for digital control system with quantization effect

Wang Liang, Wang Bing-wen, Guo Yi-Ping

(Department of Control Science and Control Engineering, Huazhong University of Science and Technology, WuHan, 430074, P.R.China) E-mail: wl@smail.hust.edu.cn

**[Abstract]** Quantization problem in digital control system have attracted more and more attention in these years. Normally, a quantized variable is regarded as a perturbed copy of the unquantized variable in the research of quantization effect, but this model has shown many obvious disadvantages in control system analysis and design process. In this paper, we give a new model for quantization based 'cell mapping' concept. This cell model could clearly describe the global dynamics of quantized digital system. Then some important characteristics of control system like controllability are analyzed by this model. The finite precision control design method based on cell concept is also presented.
**[Keywords]** Quantization effect, digital controller, Cell mapping

## 1. Introduction

Now almost all the advanced control systems are digital control system. Its infinite precision design theory and application have been well studied for a long time. However, for the finite word length limitation resulting from actual devices, modules or subsystems, the system performance will always deviate from the ideal performance in practice. So the quantization effects need to be considered when a digital controller is designed and implemented, especially for the fixed-point system. The new motivation for considering quantization in feedback control systems is the recent boom of interest in networked control systems. In these systems, the output measurements to be used for feedback are transmitted via a digital communication channel.

The term "quantization" refers to the restriction of a variable to a discrete set of values rather than a continuous set of values. There are two main disparate research directions in this area.

Some researches are concerned with the finite word length (FWL) effects in digital controller. For example, the calculations performed by digital processors have round off errors. FWL effects in memory could result in the controller parameter perturbations. The main approach of FWL controller study is to obtain the optimal realization of a given FWL controller to minimize the FWL effects according to some special measures, such as the pole sensitivity measure [1,2] and complex stability radius measure [3]. Most important results for this research could be found in [4]. Now more and more control systems are designed based on PC platform. Even many cheap digital processors has enough word length, so lately, there was almost no new progress in this area.

Another research cares about the quantization effect in the interface of digital controller and plant. For example, the quantization arises due to the use of A/D and D/A converters. Now many sensors only provide the digital signals. For the basic work theory limitation, it's also difficult to improve the precision of some sensors. Just like the level sensors for measuring the height of a fluid in a vessel, encoders for determining the angular position of induction motors or (magnetic/optic) disc drives, transportation systems, where the position of a vehicle is only known when certain markers have passed [5]. So they could only provide the quantization signals. In many modern factory control system like Fieldbus Control System (FCS), all controllers and plants are interconnected over a digital communication network, making quantization essential in order to transmit information among different parts of the system [6,7]. Driven by these actual requirements, how to apply the quantized sampled signals and control signals to achieve the control performance is becoming the main topic of quantization research. In this paper, we only consider this kind of quantization effect.

Before 1990's, almost all the researches regard a quantized variable as a perturbed copy of the unquantized variable. An ordinary stochastic model for this perturbation is a white random process having uniform probability density function. The effect of quantization on resulting performance is then analyzed utilizing a bound on the perturbation introduced by the quantizer. This approach is well justified when the required precision is easily achievable and the cost of the resulting implementation is reasonable. But more and more disadvantages of this statistical model have been found in these years [8].

First, some important assumptions in this model are not very appropriate. This statistical model comes from digital filter research. In actual feedback control systems, signal trend to be more deterministic and exhibit stronger correlation over time. The statistical modeling approach doesn't always predict certain behaviors well, especially for nonlinear system.

Second, this model could only give some statistical results like Signal-to-Noise ratio and bounds of perturbation. It's difficult to analyzing the impact of quantization to controllability, stability, robust and other important performance of control system, to say nothing of developing finite-precision design method of control system based on this model.

Thirdly, even many statistical results obtained by this model are not believable. Using modern ergodic theory, the paper [9] show the long-term behavior one sees in a feedback control system, although amenable to probabilistic analysis, can be quite differ from what a "white noise" model for quantization "errors" would predict. When ordinary 'linear' feedback of quantized state measurements is applied, the resulting closed-loop system behaves chaotically.

So the way in which quantization is dealt with in feedback control system has experienced a striking change in recent years. Several researchers have obtained some interesting results by treating quantization in unconventional ways. For example, some people treat quantization characteristic as nonlinearity. The discrete event model is also applied to investigate the quantization effect. But the most promising one is the information coder model proposed by Curry and Delcamps [9]. This approach is to regard a quantized variable as a partial observation of the unquantized variable. It means that a quantized variable provides information on a rage of values that the unquantized variable may take, rather than one specific value. These works have paved the way for the most recent approach to quantization.

Appling this new model, numerous works explicitly deal with quantization while focusing on stabilization of control system, especially for networked control system. Within there works, we can distinguish between the ones where the quantization strategy is dynamic and time-varying. If the quantization strategy is dynamic and time varying, then quantization having a finite number of levels may be employed to achieve asymptotic stabilization [10,11]. On the other hand, if the quantization strategy is fixed and static, employing a quantizer with a finite number levels can only yield local practical stability [12,13].Quantizer with a finite number of levels have greater practical significance than quantizers with an infinite number of levels. However, the latter quantizers have been very useful for proving many important results in networked control.

Throughout this paper, we will regard a quantizer as a fixed and static component of the system and will analyze different aspects of quantization in feedback control system. Now the information coder model is mainly applied to find a quantized feedback control law that stabilizes the system. We extend this model by cell mapping concept [14]. This extension could give more explicit description for many quantization effects. Based on this new model, we could also precisely describe the stability, controllability, robust and other characteristics of quantized feedback control system. The finite precision design method could also be developed by this model.

This paper is organized as follows. In the second section, we introduce the cell mapping concept to describe the dynamics of quantized digital control system. The following section shows how to analyze the some important performances of control system by ell mapping method. In the fourth section, we detail the principle of finite precise optimal design approach based on cell concept. The fifth section discusses the engineering realizations of cell method. The last section is a short summary.

## 2 Cell mapping and dynamics of quantization system
### 2.1 Information coder description for quantization effect

As we said above, we will treat the quantizer as information coder. A quantizer acts as a functional that maps a real-valued function into a piecewise constant function taking on a finite set of values. Given a positive integer M and a nonnegative real number $\Delta$, we define the quantizer $q: R \to Z$ with sensitivity $\Delta$ and saturation value M by the formula:

$$q(x) = \begin{cases} M, x > (M+1/2)\Delta \\ -M, x \leq -(M+1/2)\Delta \\ \lfloor \frac{x}{\Delta} + \frac{1}{2} \rfloor, -(M+1/2)\Delta < x < (M+1/2)\Delta \end{cases}, \text{ here } \lfloor x \rfloor := \max\{k \in Z, k < x\}$$

Then on the interval $((k-1/2)\Delta, (k+1/2)\Delta]$ of length $\Delta$, where $k \in Z$ and $-M < k < M$, the function q takes on the value k. Suppose that we have n quantizers $q_i: R \to Z$ with sensitivities $\Delta_i$ and the same saturation value $M (i = 1, \cdots n)$. We define the quantizer $q: R^n \to Z^n$ with sensitivity $(\Delta_1, \cdots, \Delta_n)$ And saturation value M as follows: $q(x) = (q_1(x_1), \cdots q_n(x_n))$, where $(x_1, \cdots x_n)$ are coordinates of x relative to a fixed orthonormal basis in $R^n$. Geometrically, $R^n$ is thereby divided into a finite number of rectilinear quantization blocks, each corresponding to a fixed value of q.

The above notation is similar to many current papers using information encoder model. In quantized feedback system, it is only known which one of a fixed number of quantization blocks contains the current state at each instant of time. At the time of passage from one quantization region to another, the dynamics of the system change. So this region transition $Z^n \to Z^n$ could (should) be applied to describe the dynamics of control system. In most current research, the quantized state information $Z^n$ is only applied to construct Control Lyapunov function to stabilize LTI system. Besides the stabilization design, we want to develop systematic analysis and design methods for quantized control system using this coder information. It's just the basic idea of our paper. First, we will introduce cell mapping concept to give the precise description for this 'region transition'.

### 2.2 Brief introduction to Cell mapping

Cell mapping is a powerful computational technique for analyzing the global behavior of nonlinear dynamical systems, which was proposed by Hsu C.S in 1980's. This technique partitions a continuous state space into a finite number of disjoint cells. Normally, cells are created by dividing each axis of an n-dimensional state space into equal intervals, each denoted by an integer $z_i$. A cell z is defined as an n-tuple of intervals $(z_1, z_2, \cdots z_n)$. The union of all cells z yields an integer-valued n-dimensional cell space Z. All System states in a cell $Z_i$ are abstracted by the cell center point $Z_i^c$. This abstraction allows continuous state space trajectories to be modeled by cell trajectories. Fig.1 illustrates the approximation of the state space trajectory.

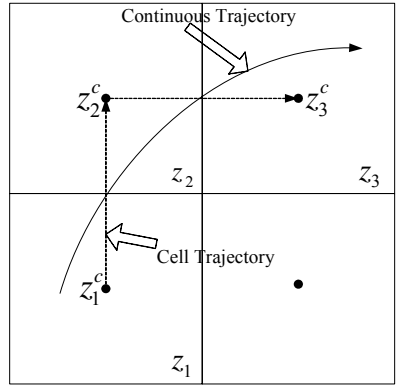

**Fig.1** Cell trajectory, the continuous trajectory could be represented by a discrete trajectory $Z_1^c \to Z_2^c \to Z_3^c$

A cell mapping is formalized as a cell state space function: C:Z→Z. The dynamic of system could be approximately described by this function. A cell map is constructed using an "unraveling algorithm" to compute cell transitions and identify all existing periodic motions and domains of attraction. The key to the unraveling algorithm is the computation of an 'image' or one-step transition cell for each cell Z within a specified 'cell processing time period'. The image cell is computed by representing cell Z by its center-point $Z^c$, and searching for the cell in Z in which the image of $Z^c$ falls. A detailed presentation of the unraveling algorithm could be found in [15].

There are several cell mapping methods. Here we use simple cell mapping (SCM) method first, in which a cell only has one "image" cell. 4 important concepts in this method are introduced as follows:

(1) A periodic motion with period K is a sequence of K distinct cells $Z_m, m=0,1,\cdots k-1$, satisfying the conditions: $Z_m = C^m(Z_0)$, $Z_m = C^k(Z_m)$

(2) An equilibrium cell $Z_e$ is a cell that maps to itself: $Z_e = C(Z_e)$. It is a periodic motion with period 1.

(3) A k-step trajectory emanating from a cell $Z_i$ is expressed as a cell sequence:

$$Z_i \to C(Z_i) \to C(C(Z_i)) \to \cdots \to C^k(Z_i)$$

(4) The r-step domain of attraction of a periodic motion is the set of all cells that are within r-steps of the periodic motion.

The main aim of cell map is to compute cell transitions and identify all existing periodic motions and domains of attraction.

**2.3 Cell description for quantization**

We could compare information coder model with cell model. In digital system, we normally divide the axis into equal piece according to the word length. Each interval is assigned a nonnegative integer. The quantized region is represent by $Z^n$. In cell method, 'cells' are created by dividing each axis of the state space into equal intervals, each denoted by an integer. A cell z is defined as an n-tuple of intervals. So each 'quantization region' in digital system could be regarded as a 'cell'.

Then for their dynamics, the abstraction of cell allows continuous state space trajectories to be described by cell transitions C:Z→Z. In digital system, we also use region transition $Z^n \to Z^n$ to describe the dynamics of system. Here the Z in cell mapping and region $Z^n$ is the same things. The regions transition is just the cell transition.

So the dynamics of quantized system could be described by cell mapping. Most theories and methods in cell mapping could all be easily applied to quantized control system. We just need construct "cells" according to the A/D word length of digital system. It's the basic principle of this paper.

Here we use a simple first-order system to illustrate the cell description of digital system:

$x(k) = 0.625 x(k-1)$, here $x(0) = 0.87$, $x \in [0,1]$. Assume the word length of A/D and D/A is 3, select round off mode [Fig.2].

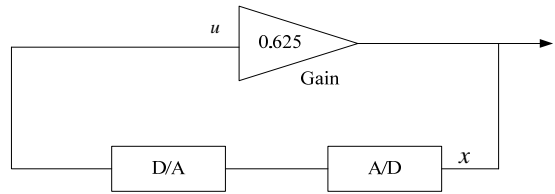

**Fig.2** A first order digital system

If we don't consider the quantization effect: $x(k) = 0.87(0.625)^n \delta(k)$. When $k \to \infty, x(k) \to 0$

Then considering quantization effect (in binary form):

$$n = 0, x(0) = [0.100 \times x(-1)]_R = 0.111$$
$$n = 1, x(1) = [0.101 \times 0.111]_R = 0.100$$
$$n = 2, x(2) = [0.101 \times x(1)]_R = 0.011$$
$$n = 3, x(3) = [0.101 \times x(2)]_R = 0.010$$
$$n = 4, x(4) = [0.101 \times x(3)]_R = 0.001$$
$$n = 5, x(4) = [0.101 \times x(3)]_R = 0.001$$
$$\vdots$$

We could get the new series of x(n): 0.875,0.5,0.375, 0.25,0.125,0.125,0.125,… The x(k) will stay at 0.125,but not 0. It's just famous zero-input limit cycles [16]. Formerly, This result could be explained as follows: When n>3, the system is equal to $x(k) = x(k-1)$. So its transfer function: $H(z) = \dfrac{1}{1-z^{-1}}$. Its pole point moves from $z = 0.625$ to $z = 1$. So the system becomes an oscillation system.

We could find this explanation relies on the input signal. If the input changed, we can't promise the system has the same behavior. Then we use cell mapping method to analyze this system:

(1) Construct cells. Because the word length is 3 in this system, we can divide the interval [0,1] into $2^3 = 8$ equal pieces and mark them as $1,2,\cdots,7,8$ respectively, which is shown in Fig.3. If select float mode, the cell will have different sizes.

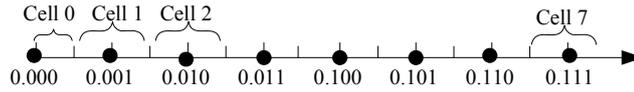

**Fig.3.** Construct cells

(2) Calculate the one-step transition of all the cells, which means getting the "image cell" of cell $1,2,\cdots,7,8$ respectively. Here center point of interval is used to represent the whole interval to calculate the cell transition. We could get one-step transition vector: [1  2  2  3  3  4  5  5].

(3) Decide the periodic motion and corresponds attraction domain. The dynamics of this cell map system is shown in Fig.4:

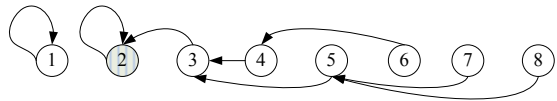

**Fig.4** Cell dynamics of a zero-input limit cycle system

We could find the cell $Z_1$, $Z_2$ is equilibrium cell. All the other cells are included in the 3-step domain of attraction of $Z_2$. That also means most input signals will reach cell $Z_2$ (corresponds to 0.125) in 3 steps. This explanation is more explicit and precise than pole point shift explanation.

**2.4 Cell description for a more complex system**

Here we also give a two dimension regulator control system to illustrate the Cell description.

$$x(k+1) = \begin{bmatrix} 0 & 1 \\ -1 & 1 \end{bmatrix} x(k) + \begin{bmatrix} 0 \\ 1 \end{bmatrix} u(k), \quad x,u \in [0,1]$$

Performance index：

$$J = \frac{1}{2}\sum_{k=0}^{\infty}[x_1^2(k) + u^2(k)]$$

It's a standard LQR problem. We could get its optimal control law according to Riccati equation. The linear state feedback vector:

$$F = [0.654 \quad -0.486],$$

So:

$$x(k+1) = \begin{bmatrix} 0 & 1 \\ -1 & 1 \end{bmatrix} x(k) + \begin{bmatrix} 0 \\ 1 \end{bmatrix}[0.654 \quad -0.486]x(k) = \begin{bmatrix} 0 & 1 \\ -0.346 & 0.514 \end{bmatrix} x(k)$$

Here we use 4 bit A/D, so there are $2^4 \times 2^4 = 256$ cells. The cell description for this system is shown in Fig.5:

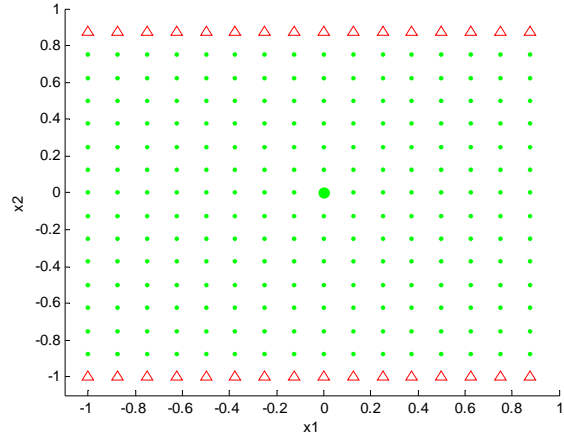

**Fig.5** cell description for dynamics of a two dimension system, the cell represented by red triangle will transit to sink cell, the other dot cell will transit to equilibrium cell

In Fig.5, we could find Cell maps contain two groups. One group contains the sink cells. The other group contains cells that converge to their equilibrium cell.

### 2.5 Global Cell mapping description for quantized system

In simple cell mapping adopted above, one cell has only one image cell. We could find some possible errors in this kind of cell description for quantized feedback control system:

For example, a LTI feedback system $X(k+1) = AX(k) + Bu[X(k)]$, the $X(k)$ in $u[X(k)]$ is quantized value and the AX(x) is the 'precise' value. But in the calculation of one step transition, we also treat the $X(k)$ in $AX(k)$ as quantized variable:

The real condition: $X(k+1) = AX(k) + Bu[q(X(k))]$

In simple cell calculation: $X(k+1) = Aq(X(k)) + Bu[q(X(k))]$

So for the same observed state $q(X(k))$, the state of next step $X(k+1)$ may be different. Here to give the 'precise' description for real dynamics of quantized system, one cell should have different image cells. The Generalized cell mapping (GCM) technology could meet this requirement [17].

Under SCM each regular cell is mapped onto a single image cell. In reality, the image of a cell will be given by some bounded region that in general may cover more than a single cell. This means that for each regular cell the number of image cells should not be restricted to one and should not be fixed. This idea supports the GCM concept. Under GCM, a regular cell can have several images, each with a fraction of total probability. The GCM method provides a probabilistic description of the time evolution of the system response.

The GCM is calculated as follows: Each cell is mapped forward one time step and it is accounted for to which other cells parts of that cell are mapped. The transition probability $w_{ij}$ for a transition from cell $Z_j$ to cell $Z_i$ is calculate as that portion of its volume (given by the Lebesgue measure u) that is mapped to $Z_i$:

$$w_{ij} = \frac{u(f^{-1}(Z_i) \cap Z_j)}{u(Z_i)}$$

The sink cell plays a special role: transitions from the sink cell back to the regular cells are not admitted, therefore, this cell is called coffin cell in stochastic dynamics. We set

$$w_{i0} = \begin{cases} 1, & \text{if } i = 0 \\ 0, & \text{if } i \neq 0 \end{cases}$$

The matrix $W = w_{ij}, i, j = 0, \cdots, N$ is column stochastic, I.e.,

$$\sum_{i=0}^{N} w_{ij} = 1, \quad w_{ij} \geq 0$$

It is important to note that the generalized cell mapping can in general only be approximated numerically. So far the sampling method or Monte-Carlo methods are used for this approximation [18].

For the one dimension system above, its one step transition matrix is P: (There is no cell transiting to sink cell):

$$P = \begin{array}{c} \\ z_1 \\ z_2 \\ z_3 \\ z_4 \\ z_5 \\ z_6 \\ z_7 \\ z_8 \end{array} \begin{array}{cccccccc} z_1 & z_2 & z_3 & z_4 & z_5 & z_6 & z_7 & z_8 \\ 0.8125 & 0.1875 & 0 & 0 & 0 & 0 & 0 & 0 \\ 0 & 1 & 0 & 0 & 0 & 0 & 0 & 0 \\ 0 & 0.4063 & 0.5938 & 0 & 0 & 0 & 0 & 0 \\ 0 & 0 & 1 & 0 & 0 & 0 & 0 & 0 \\ 0 & 0 & 0 & 1 & 0 & 0 & 0 & 0 \\ 0 & 0 & 0 & 0.5938 & 0.4063 & 0 & 0 & 0 \\ 0 & 0 & 0 & 0 & 1 & 0 & 0 & 0 \\ 0 & 0 & 0 & 0 & 0.1875 & 0.8125 & 0 & 0 \end{array}.$$

This matrix is could also be described as Fig.6:

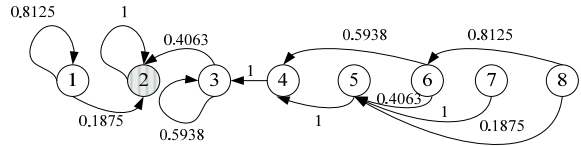

Fig.6 the General cell mapping description for the dynamics of a zero-input limit cycles system

To give the more precise description for that two dimension system above, we could write it as:

$$x(k+1) = \begin{bmatrix} 0 & 1 \\ -1 & 1 \end{bmatrix} x(k) + q\left( \begin{bmatrix} 0 \\ 1 \end{bmatrix} [0.654 \quad -0.486] q[x(k)] \right)$$, here the q is quantization operation. Assume

the A/D is 4 bit and D/A 8 bit. The GCM description for this dynamics is shown in fig.7

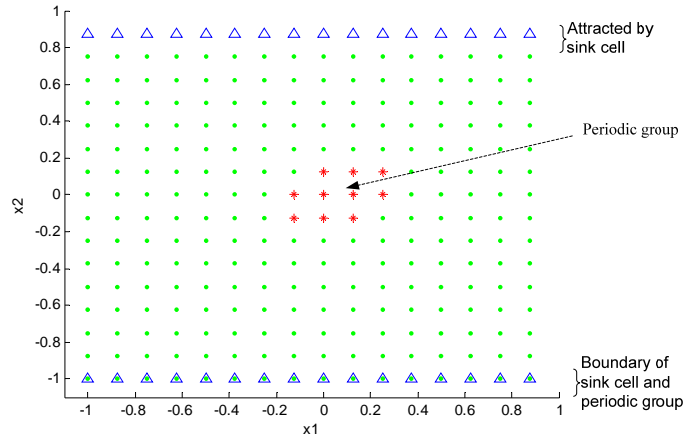

**Fig.7** the General cell mapping description for the system dynamics

In fig.7 the asterisk is the periodic group. The triangle is the cell that will transit to sink cell. The dot is cell that will drop in the period group. The combined area of dot and triangle is the boundary of sink cell and periodic group. The cell in this area may transit to sink cell or periodic group.

The SCM and GCM is the basic cell mapping technologies. Their improvement like Poincare-like cell mapping method could also be applied to analyze the dynamics of quantized system [19].

Cell mapping could give the simple and explicit description for the dynamics of digital control system with quantization effects. It's also a kind of global description, unlike some conventional methods which only analyze time responses for a few initial conditions. In the following part, we will give the further analysis for some important dynamic characteristics of digital control system by cell concept.

## 3 Evaluation the digital controller by cell mapping method
### 3.1 Stability

Stability is the base requirement for the design of a control system. Here the stability characteristic could be directly obtained from cell description. For the two dimension system above (Fig.7), we could find dot cell will drop in the periodic group at last, so these regions are stable. The cells which transit to sink could be regarded as unstable cell.

As we said above, most current researches are concerned with the stability of control system with quantization effects. They all apply the Lyapunov function to analyze stability or stabilize the system. But we know Lyapunov method is a kind of analytical method. There is still no serious discussion about whether the Lyapunov theorem is available in discrete/quantized state space system. Here the cell based stability method is more believable and clear.

### 3.2 Controllability

A linear controllable system could be defined as a system which can be steered from any state to the zero initial state. For linear time invariant digital system $\dot{x} = Ax + Bu$, we could easily judge the controllability according to rank of $[B, AB, A^2B, \cdots]$. But when we consider the quantization effect, things will become complicated. In digital system the state space is divided into a finite number of quantization regions. If the size of quantized region is very large and the sampling period is very short, the system may stay in same region in one sampling period, but not transit to another one. This could result in the distorted control sequence and system states. In some unusual situations, system may go into a limit cycle motion and the target state is not reached. This means some regions in state space may become uncontrollable for quantization effects, so a thorough study of controllability considering the quantization constraints is also desirable.

Here we meet such a stiff mission: how to determine the controllable regions with the consideration of all actual constraints: quantization of state space, time sampling and quantization of control inputs.

We could depict this problem by cell concept. For quantization effect, there are finite number of states and control inputs. For example, in a 2 dimension 8 bit A/D system, each state is divided into $2^8 = 256$ equal pieces, so there are $256 \times 256 = 65536$ cells. If selecting 2 bit D/A, there are 4 possible control values for each cell. So there are finite possible control laws for quantized system. If all possible laws can't drive one cell to equilibrium cell, this cell will be uncontrollable. Based on this idea, we could apply dynamical program method to determine the controllability of each cell.

Here we use Bang-Bang control system to detail this method, assume the A/D is 5 bit and D/A is 1bit:

$$\begin{cases} \dot{x}_1 = x_2 \\ \dot{x}_2 = u \end{cases}, \text{ or written as: } \dot{x} = Ax + Bu = \begin{bmatrix} 0 & 1 \\ 0 & 0 \end{bmatrix} x + \begin{bmatrix} 0 \\ 1 \end{bmatrix} u$$

Final state: $x_1(t_f) = 0$, $x_2(t_f) = 0$

Define a performance index as:

$$J = \int_{t_0}^{t_f} 1 dt$$

Here $x_1, x_2, u \in [-1,1]$

Some preliminary operations:

(1) Construct cells.. For 5 bit A/D sampling, $x_1, x_2 \in [-1,1]$. We divide the $x_1$ and $x_2$ into $2^5 = 32$ equal pieces. So there are $32 \times 32 = 1024$ cells, which are denoted by C.

(2) Determine the control input set U. The input set is determined by word length of D/A. Bang-Bang control system could be regarded as a 1 bit D/A system, it contains only two possible values: $U = \{-1, 1\}$.

(3) Obtain one step transition from every cell with the entire set of control U. Here assume the sampling period is 0.05s and 0.08s respectively. Because the system is given in differential equation form, the transitions are calculated by integral operation starting from the center point of each cell. All these mapping pairs are denoted by M.

Then dynamic programming is employed to identify the controllable regions. This algorithm has four steps:

(1) To find the controllable cells from the mapping data of M, we first take for all the states which can be mapped into the target set A0 in one step. This is easily done by searching through set M to locate the mapping pairs whose image states belong to A0. Obviously, these states are controllable. We marked them as A1. In Bang-bang control system, the target set A0 is the equilibrium cell.

(2) All the mapping pairs with domain states belonging to $(A0 \bigcup A1)$ will have no further relevance to our controllable regions search. We purge them from the set M and denote the remaining set by M1. Just as in step 1, we then search the states in $C \setminus (A0 \cup A1)$ that can transit into set $A0 \cup A1$ in M1 and mark them as A2. Now the set $A0 \cup A1 \cup A2$ are controllable.

(3) The pattern of search for controllable set is now clear. We need seek from $C \setminus (A0 \cup A1 \cup A2 \cup \cdots \cup Ai)$ whose image states belong to $A0 \cup A1 \cup A2 \cup \cdots \cup Ai$.

(4) The process is continued until all the controllable cells have been discovered. The end stage of the search has arrived when the number of controllable regions doesn't increase anymore.

According to this method, we could obtain the controllable regions of Bang-bang control system (Fig.8)

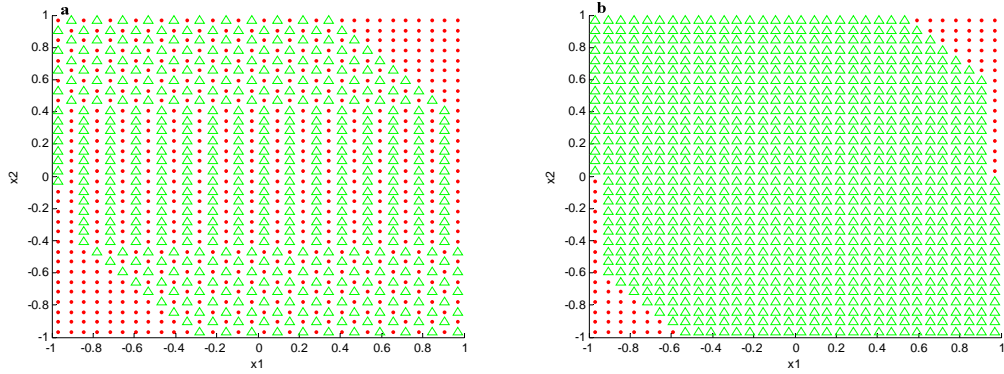

**Fig.8** Controllable state space regions of Bang-bang control system. The dot represents the uncontrollable region, the triangle is the controllable region. In figure4.a, the sampling period is 0.08s, figure.4.b is 0.05s.

In figure.8.a, there are 452 controllable regions, 572 uncontrollable regions. In figure.8.b, 950 controllable regions, 74 uncontrollable regions.

Obviously, for given system, rank$[B, AB, A^2B, \cdots]$=2, it is controllable. But our discussion above shows some state regions may become uncontrollable for quantization effect. Because the quantization is a complex nonlinear operation, it's difficult to give the careful theoretical study for the relation of controllability and quantization by conventional control theory. The GCM may be used for this research topic in the future.

**3.3 Robustness**

Robustness is the ability of a controller to maintain good performance characteristics despite parameter variations outside the design ranges. While it has been the subject of much research in classical and modern control theory, mathematical investigations of quantized digital system robustness are difficult for the state space quantization. For example, if the size of quantized region is very large, the parameter variations of plant may not affect the value of observed state.

As a computational technique, the cell map is well suited to evaluating the robustness of quantized controllers. In fact, it has been applied to analyze the robustness of fuzzy system [20]. We give the trajectory-based measures for robustness of quantized system referring to paper [21]. To evaluate controller robustness, it is best to consider how much each cell trajectory is affected by parameter variations. So for SCM, we could use the changes of one step transition vectors constructed with different system parameter values to evaluate the robustness.

Here we apply number of modified cell to express the changes of transition vector. Still for that one dimension system $x(k) = 0.625x(k-1)$, its one step transition vector [1  2  2  3  3  4  5  5]. If its parameter is 0.90, the corresponding transition vector is [1 2 3 4 5 5 6 7]. So the number of modified cell is 6. Here assume the parameter 0.625 changed between [0.525, 0.725], the modified cell number is shown in Fig.9:

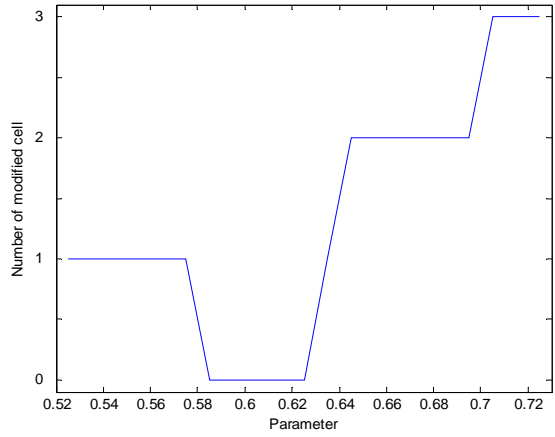

**Fig.9** Number of modified cell

Then we apply another measure, 'Number of controllable cells' to evaluate the robust of system. The number of controllable Cells is a quantitative measure of controllability. It is computed by constructing a cell map for the system plus controller and counting the number of cells in the controllable cell group. Then we examine the variances of this performance measures obtained from cell maps constructed with different system parameter values.

Still for that two dimension regulator control system:

$$x(k+1) = Ax(k) + Bu(k) = \begin{bmatrix} 0 & 1 \\ -1 & 1 \end{bmatrix} x(k) + \begin{bmatrix} 0 \\ 1 \end{bmatrix} u(k) \text{, 4 bit A/D, } 2^4 \times 2^4 = 256 \text{ cells}$$

Then add a parameter disturbance $\Delta A = \begin{bmatrix} 0 & a \\ 0 & 0 \end{bmatrix}$ to A, here $a \in [0, 0.5]$, a 50% variation. The 'Number of controllable cells' is shown in Fig.10.

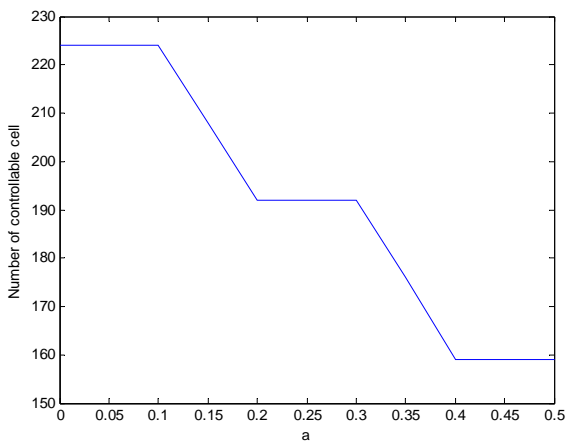

**Fig.10** Number of controllable cells

The number of controllable cells consistently drops from a high of 224 cells (87.5% controllability over the region of interest) for the design value to a low of 159 cells (62.1% controllability). Obviously, this is not a very good controllability for 50% parameter variation.

The transition matrix of GCM could give the more precise measure for robustness. Some other cell based robustness evaluation methods like 'Actual trajectory error' could also be applied to quantized system [21]. Because Perturbation in plant may not affect the observed state, we may also develop some coarse quantization design method to improve the robustness in the further research.

**4 Finite-precision design of digital control system**

The most common method to design a digital control system is to disregard the quantization in the first stage. After the controller was designed, we apply some optimization control methods to minimize quantization effects. We could find the obvious shortages in this process. When designing the controller, we normally set some optimal aims. But after being quantized, we can't ensure the system could still reach the former aim. The quantization minimizing operation make this problem more complicate. Maybe we could consider the quantization in the design process, but most optimal methods are based on analytic theory and not efficient to deal with the discrete/quantized variable. So as far as we know, there was still no deep research for finite precision optimal design problem.

As we said in first paragraph, when ordinary 'linear' feedback of quantized state measurements is applied, the resulting closed-loop system behaves chaotically. Here we use a simple example to describe this chaos:

A given system:

$$\begin{cases} \dot{x}_1 = x_2 \\ \dot{x}_2 = -x_1 + u \end{cases}$$

$$x_1(0) = 0.5, x_2(0) = 0.5$$

Assume the sampling time is 0.1s, this system is discretized as:

$$\begin{cases} x_1(n+1) = 0.9950 x_1(n) + 0.0998 x_2(n) + 0.0050 u(n) \\ x_2(n+1) = -0.0998 x_1(n) + 0.9950 x_2(n) + 0.0998 u(n) \end{cases}$$

Define a performance index as:

$$J = \sum_{k=1}^{\infty} [x_1^2(k) + u^2(k)], \quad x_1(\infty) = 0, x_2(\infty) = 0$$

It's a standard Linear-quadratic regulator design problem. Applying the Riccati equation, we could get the state-feedback law $u = -(0.3513 x_1 + 0.8886 x_2)$. The whole control system is shown in Fig.11. Here

$$A = \begin{bmatrix} 0.9950 & 0.0998 \\ -0.0998 & 0.9950 \end{bmatrix}, \quad B = \begin{bmatrix} 0.0050 \\ 0.0998 \end{bmatrix}, \quad K = \begin{bmatrix} 0.3513 \\ 0.8886 \end{bmatrix}$$

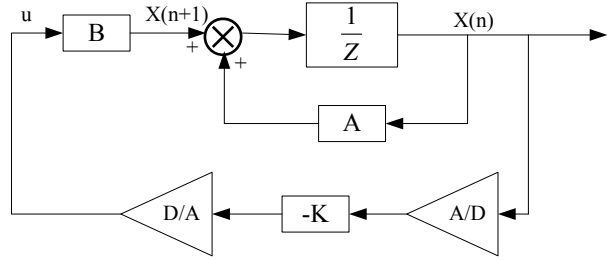

**Fig. 11** Whole feedback control system

Then we consider the quantization effect in this system. Now there has been some research for quantized feedback problem in control system. Here assume the word length of D/A is 4 bit, A/D is 8 bit, range are all [-1, 1]. The ideal and actual response of this system is shown in Fig.12:

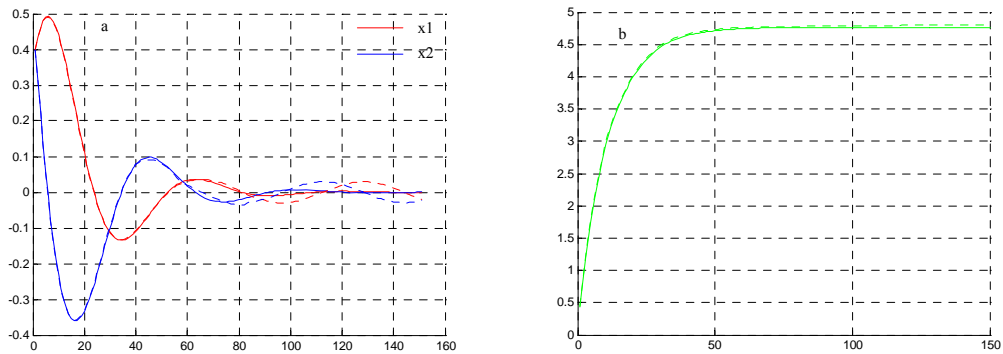

**Fig.12.a** State of given system. b. Cost of system. (Real line is precise system; the dashed is quantized input system)

We could find actual cost is a little higher than ideal system cost. But the quantized inputs can't stabilize the system. Oscillation amplitude of final states is about 0.03. We may need the complicate ergodic theory to explain this result. Here we could apply the GCM to explain this result. Its cell description is very similar to Fig.7. Most of cells will drop in the periodic group at last.

Here we meet the same difficulty in controllability analysis: how to design the control law with the consideration of state space and inputs quantization. Since the conventional analytic method is not suitable to deal with the quantized system design problem, we could consider the cell method.

In fact, Hsu has given an optimal controller design method based on cell mapping [22], though it's not special for quantized system. His method is based on the assumption of a finite set of the phase space region, sampled time intervals and admissible control inputs. The dynamic programming is employed to get the optimal feedback control law. We need only a little revision to apply Hsu's method to digital control system. To apply this method, we need construct the cell according to A/D and control input according to D/A.

The cell based optimal design method is described as follows:

(1) Construct cell. Here the word length of state variable $x_1, x_2$ is 8 and input variable u is 4. $x_1, x_2, u \in [-1,1]$.

So we divide the $x_1$ and $x_2$ into $2^8 = 256$ equal pieces. There are $256 \times 256 = 65536$ cells.

(2) Determine the control input set U. The input set is determined by word length of D/A. Its word length u is 4,

so $U = \{-1, -0.875, -0.75, -0.625 \cdots, 0, \cdots, 0.875\}$.

(3) Obtain Mappings from every cell with the entire set of control information U. For each cell, we need calculate $N_U = 16$ number of cell transitions. The related cost is also obtained. Here $x_1, u \in [-1,1]$, So the one step cost $x_1^2 + u^2 \in [0,2]$. These is still no a mature method to discretize the cost. Here we divided cost [0, 2] into 16 equal pieces and redefine the one step cost as $\{16, 17, 18, \cdots, 31\}$.

(4) Decide the only control information U for each cell. Dynamic programming is employed to identify this information. The search algorithm associates each cell with a control action that maps the cell to a cell trajectory with the optimal cost. Discrete Optimal Control Table (DOC) is obtained to represent the discrete global optimal control solution. All the optimum trajectories from every possible initial condition in the cell state space can be generated once this database is built. The detailed algorithm could be found in Hsu's original paper [22]. In our instance, there are 62453 controllable cells. A cell is said to be controllable if there exists a sequence of controls which could bring this cell to the target cell. Other 3083 cells could become controllable by increasing the search scale in state space.

(5) The optimal sequence of control is readily obtainable from DOC. Because DOC stores the singe control input for every cell, in operation process, the system only need read the input value from DOC according to the feedback state. The cell optimal solution of our system is shown in Fig.13 :

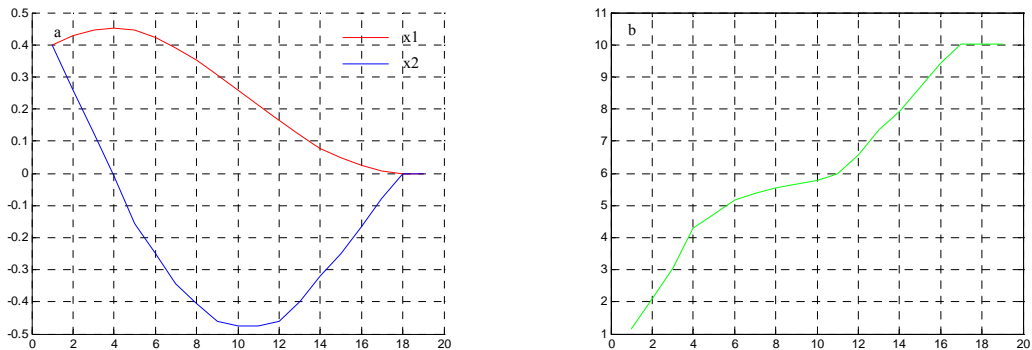

**Fig.13.a**. System state driven by cell mapping optimal control law, **b.** Cost of system

The new optimal control law is quit different to the former one. The whole cost of this system is about 10, which is higher than Riccati solution (about 4.8). But cell mapping solution could drive the system to equilibrium state, which is our main aim.

We could find obvious advantages of cell mapping based optimal design method. In most cases it is difficult to find an analytic expression of the feedback control for systems governed by ordinary differential/difference equations. But cell method is always possible to express the control in the form of a DOC table, a closed-loop feedback control laws. We should also note that cell mapping is computationally intensive, requiring considerable time and space resources, especially for precise sampling system with many variables. Now the cell mapping method is also mainly applied to deal with the fixed final state optimal design method. Some modifications and improvements have been carried out for the purpose of reducing limitations of Hsu's method [23,24].

**5 Some consideration for engineering realization of cell method**

Normally, the control laws are realized by software or special hardware. We could also give two kinds of engineering realization:

(1) Software realization. We could store the DOC in a consecutive segment of memory. For example, the DOC of 2 dimension (system state $x_1, x_2$) 4 bit A/D system could be stored in address 0000 0000—1111 1111 of memory, the high 4 bit address corresponds to $x_1$, the low 4 bit corresponds to $x_2$. If the state signal is $x_1 = 0001$ and $x_2 = 1000$, the address 0001 1000 just store the control input of this state. If there is no state signal quantization, we need design a special program to judge which cell this state belongs to. But here the state signal from sensor could be directly applied, no additional conversion needed.

(2) Hardware realization. Based on the distinct characteristics of DOC, we could develop a special hardware realization method. Here we use only memory to complete DOC operation, which is very similar to the software realization. We should have basic knowledge about memory like ROM to understand this realization. Here quantized state signals are directly connected with address lines of memory. The date lines of memory connect the D/A. We still use that 2 dimension (system state $x_1, x_2$) 4 bit A/D, 4 bit D/A system to show this method [Fig.14]:

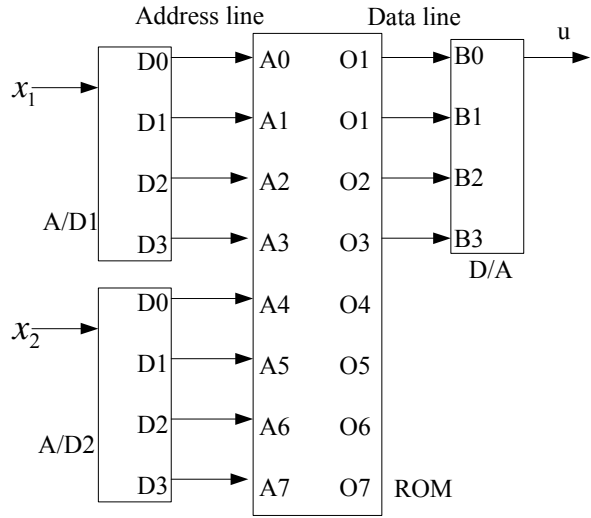

**Fig.14** DOC realization based on ROM

Its work theory is as same as software realization above. For example, $x_1$ is quantized as 0001 and $x_2 = 1000$, the address 0001 1000 is selected, this address stores the control input corresponding to this state. This 4 bit stored control input could directly export to D/A. Normally, ROM is much cheaper than special hardware like VLSI processor. This means provides a high speed and low cost realization for cell mapping method.

Here we use a practical example, the minimum-time optimal control of a DC motor system to show these methods. The equations of the DC motor in the state-space are the following:

$$\dot{x}_1 = x_2$$
$$\dot{x}_2 = -\frac{1}{\tau}x_1 + \frac{k}{\tau}u$$

Performance index: $J = \int_{t_0}^{t_f} 1 dt$, finial state: (0,0).

Where $\tau = 0.283$ s is the dominant motor time constant, $k = 0.906 rad/s^{-1}v^{-1}$ is the motor gain and $u \in [-25, 25]$ is the input drive voltage. $x_1 \in [-2.5, 2.5] rad$ is the angular position, $x_2 \in [-17, 17] rad/s^{-1}$ the angular velocity. The sampling period is 0.01 seconds. Here select 8 bit A/D and 4 bit D/A, so state space is partitioned into $2^8 \times 2^8 = 65536$ cells and control space is partitioned into $2^4 = 16$ pieces. Then we could get the DOC table according to Hsu's method. The controllable cell/region of this system is shown in Fig14.a. Assume one initial state (2, 9), its cell optimal solution is shown in Fig.15.b:

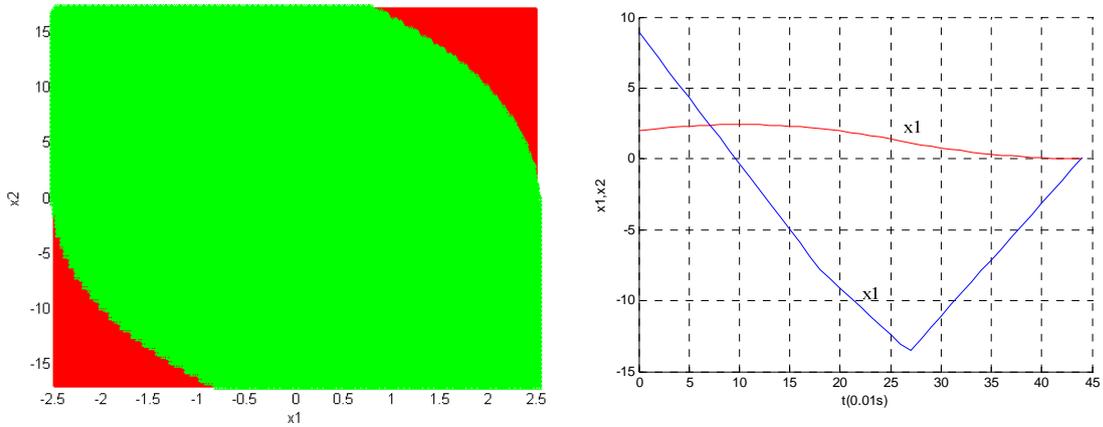

**Fig.15 a** Controllable cells of system. The green area is controllable region in state space, the red area is uncontrollable regions. **b.** System state driven by cell mapping optimal control law

We also design a hardware circuit introduced above to produce the control signal. There are $2^8 \times 2^8 = 65536$ cells in this system. So we need 64K storage to store the control information of each cell. We use a 64K*8 ROM AM27512 for this application. Part of controller circuit is shown in Fig.16.

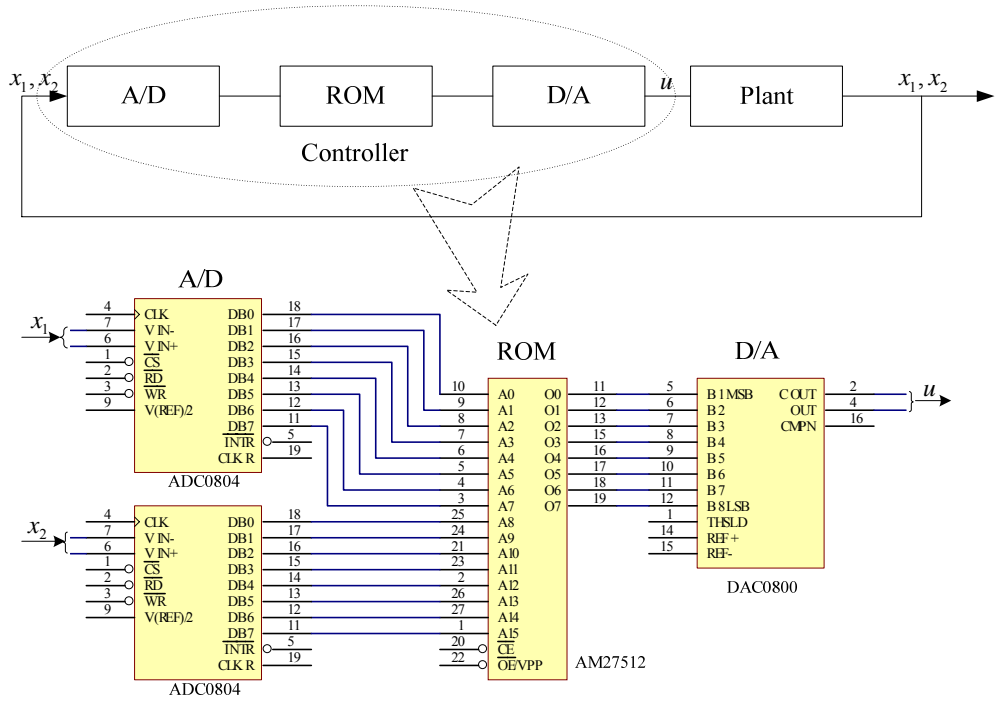

**Fig.16** Part of hardware circuit of controller

$x_1$ and $x_2$ is quantized by A/D converter ADC0804. Then the low 8 bit address lines (A0—A7) of ROM are connected with $x_1$. High 8 bit address lines (A8—A15) connect $x_2$. The data lines of ROM connect with a D/A converter DAC0800.

## 6 Conclusions

In this paper, we connect these two distinct concepts, quantization and cell mapping. Cell concept could give the clear and precise description for dynamics quantized digital control system. Here we discussed three main characteristics, stability, controllability and robustness. This relation also brings a new finite-precision design method for digital control system, which could easily deal with constraints of state and input quantization. The proposed software and hardware realization provide the useful reference for engineers who want to try cell methods in their projects.

This research features this philosophy: since the digital control system has the finite word length limitation, we'd better admit this fact and make use of it, rather than try to eliminate it. Cell mapping description emphasize discrete state space characteristic, the intrinsic and inescapable characteristic of digital system. In our opinions, it could evolve into a fundamental theory for digital control system research in the future.